\documentclass[showpacs,amsmath,amssymb,nofootinbib,prd]{revtex4}

\begin{document}

\title{Towards a wave--extraction  method for numerical relativity. V.~Extracting the Weyl scalars in the quasi-Kinnersley tetrad from spatial data}

\author{Lior M.~Burko}

\affiliation{Department of Physics, University of Alabama in Huntsville, Huntsville, Alabama 35899}

\date{January 17, 2007}

\begin{abstract}
We extract the Weyl scalars $\Psi_0$ and $\Psi_4$ in the quasi-Kinnersley tetrad by finding initially the (gauge--, tetrad--, and background--independent) transverse quasi-Kinnersley frame. This step still leaves two undetermined degrees of freedom: the ratio $|\Psi_0|/|\Psi_4|$, and one of the phases (the product $|\Psi_0|\cdot |\Psi_4|$ and the {\em sum} of the phases are determined by the so-called BB radiation scalar). The residual symmetry (``spin/boost") can be removed by gauge fixing of spin coefficients in two steps: 
First, we break the boost symmetry by requiring that $\rho$ corresponds to a global constant mass parameter that equals the ADM mass (or, equivalently in perturbation theory, that $\rho$ or $\mu$ equal their values in the no-radiation limits), thus determining the two moduli of the Weyl scalars $|\Psi_0| , |\Psi_4|$, while leaving their phases as yet undetermined. Second, we break the spin symmetry by requiring that the ratio $\pi/\tau$ gives the expected polarization state for the gravitational waves, thus determining the 
phases. Our method of gauge fixing---specifically its second step---is appropriate for cases for which the Weyl curvature is purely electric. Applying this method to Misner and Brill--Lindquist data, we explicitly find the Weyl scalars $\Psi_0$ and $\Psi_4$ perturbatively in the 
quasi-Kinnersley tetrad.

\end{abstract}


\maketitle

\section{Introduction and Summary}

Theoretical understanding of sources for gravitational waves and their emitted waveforms is an important 
step for their detection and for gravitational-wave astronomy. Such theoretical studies are typically done as 
numerical simulations. A numerical relativity simulation comprises of three major steps: First, one needs to construct the 
initial data for the physical system of interest; Second, these initial data are evolved in time \cite{baumgarte-shapiro}, and third, the evolved data are to be interpreted, and the outgoing waveforms are to be extracted. This paper addresses the last step, specifically the problem of wave extraction. It is, however, also closely related to the first step, namely the problem of initial-data determination. Indeed, although our method is applied in this paper to initial data sets (as they are available analytically---for the specific problem of interest here), it is equally applicable also for evolution data. This paper is a continuation of its prequels \cite{beetle-burko,p1,p2,p3,p4}, and is based on the method developed therein.

The long-term goal of numerical relativity is to evolve a binary system over many orbits, starting with slow, adiabatic inspiral, going through a plunge, and culminating in a  
ringdown of the produced black hole. Despite dramatic recent advances \cite{pretorius05}, it is unlikely that current numerical relativity codes will be able to to evolve hundreds of orbits in the very near future. The alternative is to 
start the simulation with the binary being at small orbital separation, so that only few orbits are left before the plunge. 
This alternative presents the construction of initial-data sets with a fundamental problem. Specifically, it is unclear how 
to encode in the initial data the gravitational waves that would emerge from inspiral from large orbital separations in the 
past of the initial time for the simulation. In fact, different constructions of initial data for the {\em same} 
physical system lead to distinct physical solutions, with different future evolutions and different gravitational 
waveforms \cite{baumgarte-shapiro,cook}. The difference between such initial-data sets is basically the amount of spurious 
gravitational waves that are present at the initial time. Lack of control over how much ``junk" gravitational waves are 
present at the initial time implies that even if the numerical evolution were done sufficiently accurately, it would 
typically be unknown which physical system that solution represents, if the ``junk" data are strong enough. 

The last step of the numerical relativity simulation requires the extraction of the outgoing waveform. The latter---to coincide with the waves that detectors such as LIGO or LISA would measure---needs to be extracted in the Kinnersley tetrad. If the Kinnersley frame is not identified, the extracted Weyl scalar $\Psi_4$ would generally be a linear combination of all five Weyl scalars of the Kinnersley frame, and therefore include not just the information about the outgoing transverse radiative degrees of freedom, but also information about incoming modes, Coulombic data, and pure gauge (longitudinal) information. Indeed, it was shown recently that the extracted waveform changes markedly when the tetrad basis vectors are transformed \cite{p4} (see also \cite{nerozzi}), so that if the Weyl scalars are extracted in a tetrad that is not sufficiently close to the Kinnersley one, the predicted waveform would deviate significantly from the waves detected experimentally.  Identifying---from numerical relativity data---the physically relevant tetrad in which the Weyl scalars are to be extracted is therefore an important problem, and this papers---like its prequels---addresses it. 

An important element in the proposed method for wave extraction is the quasi-Kinnersely (qK) frame. A frame here  means an equivalence class of tetrads (``particular set of four basis vectors") that are related to each other by type III (``spin/boost") rotations  (and exchange) transformations \cite{p2}. The quasi-Kinnerley frame is a frame that evolves continuously into the asymptotic Kinnersley frame, as the simulated (generically Petrov type--I) spacetime settles down to quiescence (generically Petrov type--D spacetime).  A particular subset of all quasi-Kinnersley frames is the {\em transverse} quasi-Kinnersley frame, defined by the requirement that it is a quasi-Kinnersley frame for which the Weyl scalars $\Psi_1$ and $\Psi_3$ vanish. Transverse frames come in threefold in type--I spacetimes, the transverse quasi-Kinnersley frame being one of these three.
The transverse quasi-Kinnersley frame is unique for any generic (type--I) spacetime, and hence lies its importance. In Ref.~\cite{p1} we showed how the transverse quasi-Kinnersley frame can be found from spatial data. The analysis of Ref.~\cite{p1} still leaves an unbroken residual symmetry,  namely  the (continuous) spin/boost rotations (and the (discrete) exchange operation). In this paper we address the problem of breaking this residual symmetry. We do not develop a general method in this paper, that would parallel the approach of Ref.~\cite{p1}.  Instead, we take the approach of addressing a particular (simple, but not too simple) spacetime, and attempting to {\em ad hoc} break the residual symmetry. We believe that this paper sheds light on the general symmetry breaking problem, as we discuss below.

Specifically, in this paper we consider two initial data sets for the time-symmetric (momentarily stationary) head-on collision of two non-rotating black holes, namely Misner \cite{misner} and Brill--Lindquist \cite{brill-lindquist} data. These initial-data sets differ in the 
topological properties of the geometry extended through the black holes' throats. For an external observer, the two 
initial-data sets differ in the amount of distortion that each black hole suffers because of the presence of the other black 
hole. Time symmetric black hole binaries were popular in the early days of numerical relativity, and 
their gravitational-wave contents have been compared using other methods \cite{anninos,abrahams}. In this paper, we revisit 
this question using a new method, that we believe to be robust: some questions  get answers that are 
background--, coordinate--, and 
tetrad--independent, in addition to being fully calculable from pure spatial data. Other questions requires gauge fixing in order to be answered, but this gauge fixing can be done in a robust and generic way for at least some problems. 

The benefit of time-symmetric black hole 
binary data is that they allow for analytic (perturbative) solutions, at least for small black hole separations. Specifically, the magnetic part of the Weyl curvature vanishes for time-symmetric data. Not only does this property simplify the derivation of the solution considerably, it also facilitates the gauge fixing that is required in order to find the quasi-Kinnersley tetrad. Unlike earlier works, we can answer the question of which initial data set has more ``junk waves" without actually evolving the data: our method allows us to find the energy flux in gravitational waves using pure spatial data.

For either initial data set, we obtain the solution for the Weyl scalars $\Psi_0$ and $\Psi_4$ using a combination of gauge invariant and gauge fixing techniques. Specifically, we first find the transverse 
quasi-Kinnersley frame, which is a family of (null) transverse tetrads connected by a spin/boost rotation. The transverse quasi-Kinnersley frame can be found in a completely gauge-, tetrad-, and background-independent method from pure spatial data as are commonly used in numerical relativity. The residual spin/boost symmetry implies that there are four degrees of freedom to be determined: the two moduli and the two phases of the Weyl scalars $\Psi_0$ and $\Psi_4$. However, already at this gauge-invariant level two of these four degrees of freedom are determined: we determine the {\em product} of the two moduli and the {\em sum} of the two phases by finding the so-called Beetle--Burko (BB) radiation scalar $\xi$ \cite{beetle-burko}, which is invariant under spin/boost rotations. Recall, that in a transverse frame $\xi=\Psi_0\,\Psi_4$, so that the phase of the BB scalar $\xi$ equals the sum of the phases of the two Weyl scalars, and its modulus equals the product of the moduli of $\Psi_0$ and $\Psi_4$, when the latter are in a transverse (quasi--Kinnersley) frame. 

We break the remaining spin/boost symmetry in two steps: First, we fix the boost degree of freedom by requiring that the expansion (or spin coefficient $\rho$) corresponds to a global mass parameter that equals the system's ADM mass \cite{p6}. This method is equivalent for the specific problem of interest to the requirement that $\rho$ (or, equivalently $\mu$)  
coincides with the no radiation limit (in which the two black holes degenerate into one). While this gauge fixing is problem adapted, we believe it---or an adaptation thereof---is general enough for most applications in numerical relativity simulations. Specifically, we require that the spin coefficient $\rho$ (or, equivalently, the expansion $\theta$) coincides with that of a Schwarzschild black hole when the proper distance along the spacelike geodesic connecting the throats of the two black holes vanishes. This method is applicable also to spacetimes that include perturbed rotating black holes. This gauge fixing determines the two moduli $|\Psi_0|$ and $|\Psi_4|$ completely.  

The remaining degree of freedom can also be determined by gauge fixing, at least for the specific problem of interest here, although not without additional information. Specifically, we argue that the spin rotation can be fixed by choosing the {\em ratio} of the spin coefficients 
$\pi$ (which below we denote by $\varpi$) and $\tau$. In principle, each of these spin coefficients has a non-zero spin weight and vanishing boost weight, and could be used to fix the remaining degree of freedom. However, both have zeros, where  spin rotations leave them vanishing. Consequently, gauge fixing is not unambiguous. However, their ratio exists everywhere (for the initial data sets of interest in this paper), and has vanishing boost weight and non-zero spin weight, so that its use is vary natural, and completely determines the Weyl scalars $\Psi_0$ and $\Psi_4$. Specifically, we extend results from point particle motion and expect that only the $h_+$ polarization state is excited significantly. Indeed, most studies of point particle motion near black holes show that the $h_{\times}$ polarization state is extremely hard to excite \cite{berti-06}. As the motion of the two black holes in the problem of interest is radial, indeed no excitation of the $h_{\times}$ polarization state is expected. We therefore choose the gauge fixing so that the generated waves include the $h_+$ polarization state only.

Our proposed gauge-fixing requires knowledge of the polarization state of the emitted gravitational waves. While the total emitted energy in gravitational waves is independent of this second step of gauge fixing (and therefore the flux of energy can be found without first breaking it), its distribution between the $h_+$ and $h_{\times}$ polarization states is not. We argue that a physical assumption on the polarization of the waves needs to be made. Our method of gauge fixing, specifically its second step, is not specific to the head-on collision case. We expect it to be appropriate for cases for which the Weyl curvature is purely electric. When the magnetic part does not vanish, $\pi$ and $\tau$ may have non-coinciding zeros, thus hindering the use of their ratio. The breaking of the residual spin symmetry when the Weyl curvature has a non-vanishing magnetic part awaits further consideration.

The organization of this Paper is as follows: In Section \ref{s2} we review the construction of the quasi-Kinnersley frame. We then construct this frame---and break first the boost and then the spin symmetry to find the quasi-Kinnersley tetrad---for Misner (Section \ref{s3}) and Brill--Lindquist (Section \ref{s4}) initial data sets. Finally, we compare our results for Misner and Brill--Lindquist initial data in Section \ref{s5}.  

\section{Construction of the quasi-Kinnersley frame}\label{s2}

In this Section we outline the calculation of the quasi-Kinnersley frame from spatial data. See Ref.~\cite{p1} for more detail. 
In a $3+1$ decomposition, the gravitational 
field is expressed in
terms of the spatial metric $\gamma_{ij}$ and the extrinsic curvature
$K_{ij}$.  From these, the electric and magnetic components $E_{ij}$ and
$B_{ij}$ of the Weyl tensor can be computed from
\begin{eqnarray}
{\cal E}_{ij} & = & R_{ij} + K K_{ij} - K_{ik} K_{j}^{~k} - 4 \pi S_{ij} \\
E_{ij} & = & {\cal E}_{ij} - \frac{1}{3} \gamma_{ij} \gamma^{kl} {\cal E}_{kl}
\label{E}
\end{eqnarray}
where $S_{ij} = \gamma_i^{~a} \gamma_{j}^{~b} T_{ab}$ is the spatial
projection of the stress energy tensor $T_{ab}$, and
\begin{eqnarray}
{\cal B}_{ij} & = & - \epsilon_i^{~kl} \nabla_k K_{lj }\\
B_{ij} & = & {\cal B}_{(ij)}.  \label{B}
\end{eqnarray}
Here, $\epsilon_{ijk}:=\tau^m\epsilon_{mijk}$ is the intrinsic spatial volume element, where $\tau^m$ is the unit future-directed normal to the hypersurface, and $\epsilon_{mijk}$ is the spacetime volume element. 

By construction, $E_{ij}$ and $B_{ij}$ are both symmetric and
traceless.  We then define the complex tensor
\begin{equation}
C^i_{~j} = E^i_{~j} + i B^i_{~j},
\end{equation}
from which the scalar curvature invariants $I$ and $J$ can be computed
as
\begin{equation} \label{I}
I = \frac{1}{2} C^i_{~j}  C^j_{~i}
\end{equation}
and
\begin{equation} \label{J}
J = - \frac{1}{6}  C^i_{~j} C^j_{~l} C^l_{~i}.
\end{equation}
In
terms of the dimensionless Baker--Campanelli speciality index (``the $S$ invariant") \cite{bc00}
\begin{equation} \label{S}
S = 27 \frac{J^2}{I^3}
\end{equation}
the Coulomb scalar can be written as
\begin{equation} \label{chi}
\chi^{0,\pm} =
- \frac{3 J}{2 I} \frac{W_{\chi}(S)^{1/3} + W_{\chi}(S)^{-1/3}}{\sqrt{S}},
\end{equation}
where $W_{\chi}(S) = \sqrt{S} - \sqrt{S - 1}$, and the BB scalar is
\begin{equation} \label{xi}
\xi^{0,\pm} = \frac{1}{4}\, I\,
\left[2-W_{\xi}(S)^{1/3}-W_{\xi}(S)^{-1/3}\right]\, ,
\end{equation}
where $W_{\xi}(S)=2S-1+2\sqrt{S(S-1)}$.  In general, both the Coulomb
and the BB scalars admit three distinct complex roots, so that $\xi$
and $\chi$ are multiply defined.  We denote these three roots with the
superscripts ``$0,\pm$''.  This ambiguity is related to the fact that
generic (i.e., algebraically general) spacetimes admit three distinct
transverse frames, in which the Weyl scalars take different values.
The value associated with the principal branch will be denoted with
a superscript $0$. The other two branches with superscripts $\pm$,
respectively. The quasi-Kinnersley \cite{p2} value will have no superscript.
For the cases studied here, there is a clear limit in which spacetime is 
algebraically special (more specifically, is Petrov type-D). One can then 
find the eigenvector ${\hat\sigma}_0^a$ corresponding to the quasi-Kinnersley frame and the 
BB scalar in the quasi-Kinnersley frame using the methods of Refs.~\cite{p1,p3}. Specifically, the 
quasi-Kinnersley frame can be identified by choosing the eigenvector ${\hat \sigma}_0^{\alpha}$ 
of $C^{a}_{~b}$ corresponding to the eigenvalue of greatest modulus \cite{p2}.  Notably, the BB scalar is not needed explicitly in order to 
find the quasi-Kinnesley frame. The only required quantity is the eigenvector  ${\hat \sigma}_0^{\alpha}$. We emphasize that while the BB scalar is a promising tool to (partially) describe the radiative degrees of freedom, the wave extraction method is not based on finding it. Rather, it is an auxiliary quantity, which need not be found for determining the quasi-Kinnersley frame.   

Once the eigenvector ${\hat \sigma}_0^{a}$ has been found, one might like to find the elements of some tetrad in the quasi-Kinnersley frame.  This, too, can be done using only spatial, rather than space-time, data.  The real null vectors of an arbitrary tetrad project into a space-like hypersurface to define a pair of real spatial vectors.  While the normalizations of these spatial projections naturally vary if one performs a spin/boost on the tetrad, they define a pair of invariant rays in the tangent space at a point.  These rays correspond to a pair of real unit vectors $\hat\lambda^a$ and $\hat\nu^a$ which are generally completely independent of one another.  In particular, they point in exactly opposite directions only when the normal $\hat\tau^a$ to the spatial slice lies in the space-time tangent 2-plane spanned by $\ell^a$ and $n^a$.  This degenerate case, which happens when the magnetic part of the Weyl tensor vanishes, is treated separately in Ref.~\cite{p1}, and is applied below for the moment of time symmetry of Misner and Brill--Lindquist initial data sets. 
The fiducial time foliation used in numerical relativity is overwhelmingly likely be ``boosted'' relative to the 2-plane defined by a given null tetrad, and only parallel $\hat\lambda^a$ and $\hat\nu^a$ are actually disallowed.

Since $\hat\lambda^a$ and $\hat\nu^a$ determine the directions of the space-time vectors $\ell^a$ and $n^a$, they also suffice to determine the frame associated to $\hat\sigma^a$, and thus $\hat\sigma^a$ itself.  The explicit relation is 
\begin{equation}\label{sln}
	\hat\sigma^a = (1 - \hat\lambda \cdot \hat\nu)^{-1}\, 
		(\hat\nu^a - \hat\lambda^a + i\epsilon^{abc}\, \hat\lambda_b\, \hat\nu_c).
\end{equation}
Note that when $\hat\lambda^a$ and $\hat\nu^a$ are interchanged, $\hat\sigma^a$ simply changes sign, as it should since the bivector $\Sigma_{ab}$ changes sign under an exchange operation in the corresponding frame. (See Ref.~\cite{p1} for definitions and detail.) 

To invert Eq.~(\ref{sln}) and solve for $\hat\lambda^a$ and $\hat\nu^a$ given $\hat\sigma^a$, we separate $\hat\sigma^a$ into real and imaginary parts:
\begin{equation}\label{sri}
	\hat\sigma^a = x^a + iy^a.
\end{equation}
In general, neither $x^a$ nor $y^a$ is unit, but the normalization condition for $\hat\sigma^a$ demands they be orthogonal with norms satisfying $\| x \|^2 - \| y \|^2 = 1$.  Taking the electric part of Eq.~(10) of Ref.~\cite{p1} then yields
\begin{equation}\label{lns}
	\hat\lambda^a = \frac{\epsilon^{abc}\, x_b\, y_c - x^a}{\| x \|^2} \quad\mbox{and}\quad
	\hat\nu^a = \frac{\epsilon^{abc}\, x_b\, y_c + x^a}{\| x \|^2}.
\end{equation}
Finally, the elements of the most general null tetrad $(\ell^a, n^a, m^a, \bar m^a)$ in the frame associated to $\hat\sigma^a$ take the form 
\begin{equation}\label{lnproj}
	\begin{array}{@{}r@{}l@{}}
		\ell^a &{}= \frac{|c|^{-1}}{\sqrt{1 - \hat\lambda \cdot \hat\nu}}\, (\hat\tau^a + \hat\lambda^a) \\[2ex]
		n^a &{}= \frac{|c|}{\sqrt{1 - \hat\lambda \cdot \hat\nu}}\, (\hat\tau^a + \hat\nu^a) \\[2ex]
		m^a &{}= \frac{e^{i\vartheta}}{\sqrt{2}}\, 
			\frac{\sqrt{1 + \hat\lambda \cdot \hat\nu}}{\sqrt{1 - \hat\lambda \cdot \hat\nu}}\,
			(\hat\tau^a + \hat\mu^a),
	\end{array}
\end{equation}
where the \textit{complex} unit projection of $m^a$ is 
\begin{equation}\label{mproj}
	\hat\mu^a = (1 + \hat\lambda \cdot \hat\nu)^{-1}\, (\hat\lambda^a + \hat\nu^a + 
		i \epsilon^{abc}\, \hat\lambda_b\, \hat\nu_c).
\end{equation}
These vectors are the result of a spin/boost with parameter $c = |c| e^{i\vartheta}$, acting on a preferred tetrad fixed by setting $\ell^a\, \hat\tau_a = n^a\, \hat\tau_a < 0$ and $m^a\, \hat\tau_a = \bar m^a\, \hat\tau_a \le 0$.  Note that when $\hat\lambda^a$ and $\hat\nu^a$ are indeed anti-parallel, the expression for $m^a$ in Eq.~(\ref{lnproj}) is ill-defined.  However, the limit as $\hat\nu^a \to -\lambda^a$ does exist, though it does depend on how the limit is taken.  This is natural since the fixed tetrad is determined only up to spin transformations in this case. This degenerate case is treated in detail in Ref.~\cite{p1}, and is used below for the head-on collision case of two nonrotating black holes as an illustration. Notice that Eqs.~(\ref{lnproj}) are determined up to a continuous complex function $c$, that is, the basis vectors are a two (real) parameter family (``equivalence class") of tetrads. This family of tetrads is the quasi-Kinnersley frame, and the desired quasi-Kinnersley tetrad is a single member of this family of tetrads.

\section{Misner data}\label{s3}

Misner data can be written in spherical isotropic coordinates as 
\begin{equation}\label{metric}
\,d\sigma^2 = \Phi^4(R,\theta ;\mu_0)\,\left[\,dR^2+R^2\left(\,d\theta^2+\,\sin^2\theta\,d\phi^2\right)\right]
\end{equation}
with unit lapse and vanishing shift, where the conformal factor is given by 
\begin{equation}
\Phi_{\rm Misner}(R,\theta ;\mu_0)=1+2\sum_{\ell=0,2,4,\cdots }\kappa_{\ell}(\mu_0)\left(\frac{M}{R}\right)^{\ell+1}
P_{\ell}(\cos\theta )\, ,
\end{equation}
where $P_{\ell}(\cos\theta)$ are the Legendre functions. 
The parameter $\mu_0$ is related to the separation of the two holes (in a manner to be made clear below) and $M$ is the  
ADM mass of the system. The coefficients $\kappa_{\ell}(\mu_0)$ are given by
\begin{equation}
\kappa_{\ell}(\mu_0) \,:=\, \frac{1}{(4\Sigma_1)^{\ell+1}}\sum_{n=1}^{\infty}\,\frac{(\coth n\mu_0)^{\ell}}
{\sinh n\mu_0}\;\;\;\;\;\; {\rm and}\;\;\;\;\;\; \Sigma_1 \,:=\, \sum_{n=1}^{\infty}\,\frac{1}{\sinh n\mu_0}\, . 
\end{equation}
As pointed out by Anninos {\em et al.} \cite{anninos}, Misner data can represent perturbations over a single Schwarzschild 
black hole for $\mu_0\ll 1$. Repeating the approximations in Ref.~\cite{anninos}, for $\ell\ge 1$,  we assume 
that $(\coth n\mu_0)^\ell /\,\sinh n\mu_0 \approx (n\mu_0 )^\ell$ (which strictly speaking is correct for $n\mu_0\ll 1$) for 
small $\mu_0$, so that we only keep terms in summations up to $N\sim \mu_0^{-1}$. We therefore have 
\begin{equation}
\sum_{n=1}^{\infty}\,\frac{(\coth n\mu_0)^{\ell}}{\sinh n\mu_0}\approx \,\frac{\zeta (\ell+1)}{\mu_0^{\ell+1}}\;\;\;\; , 
\;\;\;\; 
\Sigma_1\approx\, \frac{|\ln\mu_0|}{\mu_0}\;\;\;\; , {\rm and} \;\;\;\;
\kappa_{\ell}(\mu_0)\approx\, \frac{\zeta (\ell+1)}{|4\ln\mu_0|^{\ell+1}}\, .
\end{equation}
Here, $\zeta$ is the Riemann zeta function. Note that $\Sigma_1=M/2$. 

The proper distance $L$ along the spacelike geodesic connecting the throats of the two black holes is
\begin{equation}
L=2\left[1+2\mu_0\sum_{n=1}^{\infty}\frac{n}{\sinh (n\mu_0)}\right]\, ,
\end{equation}
which in the limit of small separation ($\mu_0\ll 1$) is given by
\begin{equation}
L\approx\, \frac{\pi^2}{4|\ln\mu_0|}\, M\, ,
\end{equation}
so that the initial data, to hexadecapole order, can be written in this approximation as
\begin{equation}
\Phi_{\rm Misner}\approx\, 1+\frac{M}{2R}+\frac{\zeta (3)\,P_2(\cos\theta)}{2\pi^6}\,\left(\frac{L}{R}\right)^3
+\frac{\zeta (5)\,P_4(\cos\theta)}{2\pi^{10}}\, \left(\frac{L}{R}\right)^5\, ,
\end{equation}

\subsection{The quasi-Kinnersley frame}

We first find the elements of the tensor $C^i_{\;j}$ to $O(L^3)$. They are
\begin{eqnarray}
C^{R}_{\;R} &=& -2^7\,\frac{M\,R^3}{(2R+M)^6}-\frac{2^6\,\zeta(3)}{\pi^6}\,P_2(\cos\theta)\,\frac{24R^2+2MR+M^2}{(2R+M)^7}\,L^3+O(L^5) \\
C^{\theta}_{\;\theta} &=& 2^6\,\frac{M\,R^3}{(2R+M)^6}+\frac{2^5\,\zeta(3)}{\pi^6}\,\frac{2P_2(\cos\theta)(14R^2+3MR+M^2)-(2R+M)^2}{(2R+M)^7}\,L^3 +O(L^5)\\
C^{\phi}_{\;\phi} &=& 2^6\,\frac{M\,R^3}{(2R+M)^6}+\frac{2^5\,\zeta(3)}{\pi^6}\,\frac{2P_2(\cos\theta)(10R^2-MR)+(2R+M)^2}{(2R+M)^7}\,L^3+O(L^5) \\
C^{\theta}_{\;R} &=& \frac{1}{R^2}\,C^R_{\;\;\theta} \nonumber \\
      &=& -3\cdot 2^5\,\frac{\zeta(3)}{\pi^6}\,\frac{8R+M}{R\,(2R+M)^6}\,\sin\theta\,\cos\theta\,L^3+O(L^5)\, .
\end{eqnarray}
The eigenvalues of $C^i_{\;j}$ are 
$$\lambda_1=C^R_{\;R}+O(L^6)$$
$$\lambda_2=C^{\theta}_{\;\theta}+O(L^6)$$
$$\lambda_3=C^{\phi}_{\;\phi}$$
so that the eigenvalue that corresponds to the quasi-Kinnersley frame is $\lambda_1$, and the corresponding eigenvector is $v_1^{\alpha}=a\,(\,\delta^{\alpha}_R+X_M\,\delta^{\alpha}_{\theta})$, where 
\begin{equation}\label{X_M}
X_M=\frac{\zeta(3)}{2\pi^6}\,\frac{8R+M}{M\,R^4}\,\sin\theta\,\cos\theta\,L^3+O(L^5)
\end{equation}
and
\begin{equation}
a=\Phi^{-2}(1+R^2X_M^2)^{-1/2}=
\frac{4R^2}{(2R+M)^2}-\frac{8\zeta(3)}{\pi^6\,(2R+M)^3}\,P_2(\cos\theta)\,L^3+O(L^5)\, .
\end{equation}
(Notice, that $X_M$ affects $a$ only at $O(L^6)$.)

Following the method outlined in Refs.~\cite{p1,p3}, we find the BB scalar to equal
\begin{equation}\label{xi_m}
\xi_{\rm M}=\frac{2^8\cdot 3^2\,\zeta^2(3)}{\pi^{12}\,(M+2R)^{10}}\,\sin^4\theta\,L^6\,+\,
\frac{5\cdot 2^9\cdot 3\,\zeta (3)\,\zeta (5)}{\pi^{16}\,R^2\,(M+2R)^{10}}\,[7\,P_2(\cos\theta)+2]\,\sin^4\theta\,L^8
\,+\,O(L^9)
\end{equation}
and the Coulomb scalar to equal
\begin{equation}
\chi_{\rm M}= -\frac{2^6\,M\,R^3}{(M+2R)^6}\,-\,\frac{2^5\,\zeta (3)\,(24\,R^2+2\,MR+M^2)}{\pi^6\,(M+2R)^7}\,
P_2(\cos\theta)\,L^3 \,-\,
\frac{3\,\zeta (5)\,(10\,R^2+5\,M\,R+M^2)}{\pi^{10}\,R^2\,(M+2R)^7}\,P_4(\cos\theta)\,L^5\,+\,O(L^6)\, .
\end{equation}

Because this is a time symmetric configuration, the magnetic part of the Weyl tensor vanishes, from which it follows that the eigenvector of Weyl is real, as is evident from Eq.~(\ref{X_M}). This implies that the two real projections of the two real null vectors of the quasi-Kinnersley frame onto the hypersurface are anti-parallel. As $ {\hat v}_1^{\;\alpha}$ is already normalized, we take ${\hat \lambda}^{\alpha}=-{\hat \nu}^{\alpha}
= {\hat v}_1^{\;\alpha}$, and then the two null spacetime vector are
\begin{equation}\label{lM}
 \ell^{\alpha} = \frac{|c|^{-1}}{\sqrt{2}}\,\left[\,\delta^{\alpha}_t+\frac{1}{\Phi^2\sqrt{1+R^2X_M^2}}
 \left( \,\delta^{\alpha}_R+X_M\,\delta^{\alpha}_{\theta}\right)\right]
 \end{equation}
\begin{equation}\label{nM}
 n^{\alpha} = \frac{|c|}{\sqrt{2}}\,\left[\,\delta^{\alpha}_t-\frac{1}{\Phi^2\sqrt{1+R^2X_M^2}}
 \left( \,\delta^{\alpha}_R+X_M\,\delta^{\alpha}_{\theta}\right)\right]
\end{equation}
and the complex null vector is
\begin{equation}\label{mM}
m^{\alpha} = \frac{e^{i\vartheta}}{\sqrt{2}\Phi^2R}\left[\,\frac{1}{\sin\theta}\,\delta^a_{\phi}+
\frac{i}{\sqrt{1+R^2X_M^2}}\left(\,\delta^a_{\theta}-R^2X_M\,\delta^a_R\right)\right]
\end{equation}
Here, $c:=|c|\exp(i\vartheta)$ is a complex valued function on spacetime with modulus $|c|$ and argument $\vartheta$. 

The Weyl scalars in the basis (\ref{lM})--(\ref{mM}) are given by  
\begin{eqnarray}
\Psi_0 &=& |c|^{-2}e^{2i\vartheta}\,\frac{3\cdot 2^4\zeta(3)}{\pi^6}\,\frac{\,\sin^2\theta}{(2R+M)^5}\,L^3 
+ |c|^{-2}e^{2i\vartheta}\,\frac{5\cdot 2^4\zeta(5)}{\pi^{10}}\,\sin^2\theta\,[7P_2(\cos\theta) +2]\,\frac{L^5}{R^2(2R+M)^5}\nonumber \\
&-& |c|^{-2}e^{2i\vartheta}\,\frac{2^3\zeta^2(3)}{\pi^{12}}\,\sin^2\theta\,\frac{2P_2(\cos\theta )(64R^2+49MR+M^2)+(2R+M)(32R+M)}{MR^3(2R+M)^6}\,L^6+O(L^7)\\
\Psi_2 &=& -\frac{2^6\,M\,R^3}{(M+2R)^6}\,-\,\frac{2^5\,\zeta (3)\,(24\,R^2+2\,MR+M^2)}{\pi^6\,(M+2R)^7}\, P_2(\cos\theta)\,L^3 +O(L^5)\\
\Psi_4 &=& |c|^{2}e^{-2i\vartheta}\,\frac{3\cdot 2^4\zeta(3)}{\pi^6}\,\frac{\,\sin^2\theta}{(2R+M)^5}\,L^3 
+  |c|^{2}e^{-2i\vartheta}\,\frac{5\cdot 2^4\zeta(5)}{\pi^{10}}\,\sin^2\theta\,[7P_2(\cos\theta) +2]\,\frac{L^5}{R^2(2R+M)^5}\nonumber \\
&+& |c|^{2}e^{-2i\vartheta}\,\frac{2^3\zeta^2(3)}{\pi^{12}}\,\sin^2\theta\,\frac{2P_2(\cos\theta )(64R^2+49MR+M^2)+(2R+M)(32R+M)}{MR^3(2R+M)^6}\,L^6+O(L^7)\, .
\end{eqnarray}
Because this is an approximate solution, the longitudinal Weyl scalars $\Psi_1$ and $\Psi_3$ are not exactly zero, but at $O(L^5)$. Specifically, as we expressed $X_M$ to $O(L^3)$, the coefficients of  
$\Psi_1$ and $\Psi_3$ at $O(L^3)$ vanishes, and the next term, at $O(L^5)$ remains. We can nullify these (and successive) terms by keeping terms to sufficiently high order on $\Phi$ and in $X_M$.

\subsection{The qK tetrad: Breaking the boost symmetry}
We break the spin/boost symmetry in two steps: first, we break the boost symmetry while leaving the spin symmetry unbroken. At the next step we address the breaking of the spin symmetry. To break the boost symmetry we are looking for a spin coefficient with non-zero boost weight but vanishing spin weight. The two candidates are the spin coefficients $\rho$ and $\mu$. The spin coefficient
\begin{eqnarray}
\rho &:=& -m^{\alpha} {\bar m}^{\beta}\,\nabla_{\beta}\, \ell_{\alpha}\nonumber\\
&=& -2\sqrt{2}\,|c|^{-1}\,\frac{R(2R-M)}{(2R+M)^3}+O(L^3)\, .
\end{eqnarray}
Under Type III rotations $\rho$ transforms like $\ell^{\alpha}$, i.e., $\rho\to |c|^{-1}\rho$. We can therefore scale the null basis vectors $\ell^{\alpha}$ and $n^{\alpha}$ by demanding that $\rho$ equals, in the limit $L=0$ when no radiation exists, the Schwarzschild black hole value $\rho_S$. As $\rho$ is real, the twist vanishes, and $\rho$ equals the expansion $\theta$. Recall that for $L=0$, we have a single non-rotating black hole with ADM mass $M$, for which 
the spin coefficient $\rho_S=-R^{-1}(1+M/2R)^{-2}$. Requiring that for $L=0$ we get $\rho=\rho_S$, 
we find that  
\begin{equation}
|c|=\frac{1}{\sqrt{2}}\,\frac{2R-M}{2R+M}\, .
\end{equation}
Alternatively, we can recall that for $L=0$ (also for a Kerr black hole), $\Psi_2/\rho^3=M$. Under a type III rotation the combination $\Psi_2/\rho^3$ has boost weight $3$ and zero spin weight. We can therefore find a boost rotation function $|c|$ so that this combination equals a global constant. To fix the remaining multiplicative constant factor in $|c|$ we require that the global constant equals $M$.  (In practice we apply this method here in the $L\to 0$ limit.) 
Putting these expressions together, we solve for $|c|$, and recover the previous result. One can also use the spin coefficient $\mu :={\bar m}^{\alpha}m^{\beta}\,\nabla_{\beta}n_{\alpha}$, and demand that it would equal the no-radiation limit as $L\to 0$, or $\mu_S=-2R(2R-M)^2/(2R+M)^4$. This calculation recovers the value of $|c|$. 

We can now find $|\Psi_0|$ and $|\Psi_4|$ in the quasi-Kinnersley tetrad
\begin{eqnarray}
|\Psi_0| &=& \left|\,\frac{3\cdot 2^5\zeta(3)}{\pi^6}\,\sin^2\theta\,\frac{L^3}{(2R-M)^2(2R+M)^3} + \frac{5\cdot 2^5\zeta(5)}{\pi^{10}} \frac{1}{R^2(2R-M)^2(2R+M)^3}\,\sin^2\theta\,[7P_2(\cos\theta)+2]\,L^5\right. \nonumber \\
&-&\left. \frac{2^4\zeta^2(3)}{\pi^{12}}\,\sin^2\theta\,\frac{2P_2(\cos\theta )(64R^2+49MR+M^2)+(2R+M)(32R+M)}{MR^3(2R+M)^4(2R-M)^2}\,L^6+O(L^7)\,\right|
\end{eqnarray}
\begin{eqnarray}
|\Psi_4| &=& \left|\,\frac{3\cdot 2^3\zeta(3)}{\pi^6}\,\sin^2\theta\,\frac{(2R-M)^2\,L^3}{(2R+M)^7}
+\frac{5\cdot 2^3\zeta(5)}{\pi^{10}}  \frac{(2R-M)^2}{R^2(2R+M)^7}\,\sin^2\theta\,[7P_2(\cos\theta)+2]\,L^5 \right. \nonumber \\
&-&\left. \frac{2^2\zeta^2(3)}{\pi^{12}}\,\sin^2\theta\,\frac{2P_2(\cos\theta )(64R^2+49MR+M^2)+(2R+M)(32R+M)}{MR^3(2R+M)^8}(2R-M)^2\,L^6+O(L^7)\right|
\end{eqnarray}
At great distances $R\gg M$, $|\Psi_4|\sim L^3/R^5$ and $|\Psi_0|\sim L^3/R^5$.

The energy flux in gravitational waves depends on the (time domain) Weyl scalars $\Psi_0$ and $\Psi_4$ in a non-local way, as is manifest from the integral over preceding times in the expression for the energy flux in gravitational waves (at infinity)
\begin{equation}
\frac{\,dE}{\,dt}(t)=\lim_{r\to\infty}\left[\frac{1}{4\pi r^6}\int_{\omega}\,d\omega \left|\int_{-\infty}^t\,dt'\, \psi(t',r,\theta ,\phi)\right|^2\,\right]\, ,
\end{equation}
where $\psi$ is the Teukolsky function, $r$ is the geometrical radial coordinate, and $\,d\omega$ is the integration measure over the 2-sphere.  The contribution to the energy flux (at finite distances) in gravitational waves at the moment of time symmetry (``$t=0$") from that moment in time (``$t'=0$")
is given by 
\begin{eqnarray}
{\cal F}_0 &:=& \int R^2|\Psi_0|^2\,d\Omega\nonumber \\
&=& \frac{3\cdot 2^{11}\zeta^2(3)}{5\,\pi^{11}}\,\frac{L^6}{R^2(2R+M)^{2}(2R-M)^4}
-\frac{3\cdot 2^{11}\zeta^3(3)}{5\cdot 7\,\pi^{17}}\,\frac{64R^2+22MR+M^2}{MR^5(2R+M)^{3}(2R-M)^4}\,L^9\nonumber \\
&+&\frac{5\cdot 2^{11}\zeta^2(5)}{\pi^{19}}\,\frac{L^{10}}{R^6(2R+M)^2(2R-M)^4}-
\frac{2^{12}\zeta^2(3)\zeta(5)}{5\cdot 7\pi^{21}}\,\frac{160R+173M}{MR^6(2R+M)^3(2R-M)^4}\,L^{11}+O(L^{12})
\\
{\cal F}_4 &:=& \int R^2|\Psi_4|^2\,d\Omega\nonumber \\
&=& \frac{3\cdot 2^7\zeta^2(3)}{5\,\pi^{11}}\,\frac{(2R-M)^4}{R^2(2R+M)^{10}}\,L^6
-\frac{3\cdot 2^7\zeta^3(3)}{5\cdot 7\,\pi^{17}}\,\frac{(64R^2+22MR+M^2)(2R-M)^4}{MR^5(2R+M)^{11}}\,L^9 \nonumber \\
&+& \frac{5\cdot 2^7\zeta^2(5)}{\pi^{19}}\,\frac{(2R-M)^4}{R^6(2R+M)^{10}}\,L^{10}
-\frac{2^8\zeta^2(3)\zeta(5)}{5\cdot 7\pi^{21}}\,\frac{(160R+173M)(2R-M)^4}{MR^6(2R+M)^{11}}\,L^{11}+O(L^{12}) \, ,
\end{eqnarray}
where ${\cal F}_i$ is the flux in $\Psi_i$.

\subsection{The qK tetrad: Breaking the spin symmetry}
To break the spin symmetry we look for spin coefficients that have non-zero spin weight and vanishing boost weight. The two candidates are the spin coefficients $\varpi$\footnote{Conventionally, ${\bar m}^a \ell^b\,\nabla_b\, n_a$ is denoted by $\pi$. Here, to avoid confusion with the geometrical $\pi$, we change the usual notation, and denote it with $\varpi$.} and $\tau$. The spin coefficient
\begin{eqnarray}
\varpi &:=& {\bar m}^a \ell^b\,\nabla_b\, n_a\nonumber\\
&=& -2\sqrt{2}\,i\,\frac{\zeta(3)}{\pi^6}\,\sin\theta\,\cos\theta\frac{e^{-i\vartheta}}{MR^2(2R+M)}\,L^3 +O(L^5)
\end{eqnarray}
and
\begin{eqnarray}
\tau &:=& -m^a n^b\,\nabla_b\, \ell_a\nonumber\\
&=& -2\sqrt{2}\,i\,\frac{\zeta(3)}{\pi^6}\,\sin\theta\,\cos\theta\frac{e^{i\vartheta}}{MR^2(2R+M)}\,L^3 +O(L^5)\, .
\end{eqnarray}

We cannot just consider a spin rotation (i.e., a type III rotation with $|c|=1$) directly on $\varpi$ and $\tau$, because $\varpi$ and $\tau$ have zeros on the equatorial plane and at the poles. At the zeros (and numerically also in neighborhoods thereof) the rotation parameter is arbitrary. However, because both $\varpi$ and $\tau$ have zeros at the same locations, the limit of their ratio exists everywhere. We therefore consider the ratio
\begin{equation}\label{spin_ratio}
\frac{\varpi}{\tau}=e^{-2i\vartheta}\, .
\end{equation}
Notably, Eq.~(\ref{spin_ratio}) is satisfied nonperturbatively. 
Under type III rotation by angle $\varphi$, $\varpi/\tau\to e^{-2i\varphi}\varpi/\tau$ (i.e., the ratio $\varpi /\tau$ has spin weight $-2$), so that by choosing a spin rotation function $\varphi$ we can determine the ratio $\varpi/\tau$ uniquely. Choosing the function $\varphi$ we in fact determine the local polarization state of the gravitational waves in the wave zone, where $\Psi_4$ and $\Psi_0$ have a clear physical interpretation in term of outgoing and incoming waves, respectively. Notice, that once the boost degree of freedom is fixed, the total energy flux in gravitational waves is determined.  The spin degree of freedom will nevertheless move part of this flux between the two polarization states. This property suggests to us that the spin degree of freedom can only be determined based on the understanding of the polarization of the waves in the wave zone. As already discussed in the Introduction, studies on radial motion in the point particle limit have shown that the $h_{\times}$ polarization state is extremely hard to excite. Following Berti {\em et al} \cite{berti-06} we may presumably lean on the point particle results, and assume that only the $h_+$ polarization state is generated. In our case, indeed this conclusion is guaranteed by the diagonality of the metric (\ref{metric}). 

Our ability to break the spin symmetry depends on the observation that the ratio $\varpi / \tau$ is free of  zeros, which may be attributed to the vanishing of $B^i_{~j}$.\footnote{Where $B^i_{~j}\ne 0$ the real unit vectors ${\hat \lambda}^a$ and ${\hat \nu}^a$ are not co-linear, so that the limit $\varpi / \tau$ may not exist everywhere.} We assert that this method for breaking the spin symmetry by gauge fixing $\varpi / \tau$ according to the polarization state is applicable to purely electric cases for which $B^i_{~j}=0$. The question of how to break the spin symmetry for the magnetic case is not answered by the above discussion, and we hope to address this question elsewhere.
 
In practice, both $\Psi_0$ and $\Psi_4$ would be real for this case.  Choosing $\varphi=-\vartheta$ we can set $\varpi/\tau =1$. Under this rotation, $\Psi_0\to e^{2i\varphi}\Psi_0$ and $\Psi_4\to e^{-2i\varphi}\Psi_4$ , so that the new Weyl scalars equal their moduli, i.e., 
\begin{eqnarray}
\Psi_0 &=& \frac{3\cdot 2^5\zeta(3)}{\pi^6}\,\sin^2\theta\,\frac{L^3}{(2R-M)^2(2R+M)^3} + \frac{5\cdot 2^5\zeta(5)}{\pi^{10}} \frac{1}{R^2(2R-M)^2(2R+M)^3}\,\sin^2\theta\,[7P_2(\cos\theta)+2]\,L^5 \nonumber \\
&-& \frac{2^4\zeta^2(3)}{\pi^{12}}\,\sin^2\theta\,\frac{2P_2(\cos\theta )(64R^2+49MR+M^2)+(2R+M)(32R+M)}{MR^3(2R+M)^4(2R-M)^2}\,L^6+O(L^7)
\end{eqnarray}
\begin{eqnarray}
\Psi_4 &=& \frac{3\cdot 2^3\zeta(3)}{\pi^6}\,\sin^2\theta\,\frac{(2R-M)^2\,L^3}{(2R+M)^7}
+\frac{5\cdot 2^3\zeta(5)}{\pi^{10}}  \frac{(2R-M)^2}{R^2(2R+M)^7}\,\sin^2\theta\,[7P_2(\cos\theta)+2]\,L^5  \nonumber \\
&-& \frac{2^2\zeta^2(3)}{\pi^{12}}\,\sin^2\theta\,\frac{2P_2(\cos\theta )(64R^2+49MR+M^2)+(2R+M)(32R+M)}{MR^3(2R+M)^8}(2R-M)^2\,L^6+O(L^7)\, .
\end{eqnarray}
We emphasize that this breaking of the spin symmetry relies on physical intuition based on extension of the point particle results to the equal mass case. Any other choice for $\varphi$ would do just as fine in the absence of knowledge of the local polarization state of the gravitational waves. Notice, in contrast, that breaking the boost symmetry is a unique gauge fixing by virtue of the radiation free limit, and that getting the quasi-Kinnnersley frame is gauge invariant.

Finally, we can write explicitly the basis vectors for the quasi-Kinnersley tetrad:
\begin{equation}\label{lMf}
 \ell^{\alpha} = \frac{2R+M}{2R-M}\,\left[\,\delta^{\alpha}_t+\frac{1}{\Phi^2\sqrt{1+R^2X_M^2}}
 \left( \,\delta^{\alpha}_R+X_M\,\delta^{\alpha}_{\theta}\right)\right]
 \end{equation}
\begin{equation}\label{nMf}
 n^{\alpha} = \frac{2R-M}{2(2R+M)}\,\left[\,\delta^{\alpha}_t-\frac{1}{\Phi^2\sqrt{1+R^2X_M^2}}
 \left( \,\delta^{\alpha}_R+X_M\,\delta^{\alpha}_{\theta}\right)\right]
\end{equation}
\begin{equation}\label{mMf}
m^{\alpha} = \frac{1}{\sqrt{2}\Phi^2R}\left[\,\frac{1}{\sin\theta}\,\delta^a_{\phi}+
\frac{i}{\sqrt{1+R^2X_M^2}}\left(\,\delta^a_{\theta}-R^2X_M\,\delta^a_R\right)\right]\, .
\end{equation}
\begin{equation}\label{mbarMf}
{\bar m}^{\alpha} = \frac{1}{\sqrt{2}\Phi^2R}\left[\,\frac{1}{\sin\theta}\,\delta^a_{\phi}-
\frac{i}{\sqrt{1+R^2X_M^2}}\left(\,\delta^a_{\theta}-R^2X_M\,\delta^a_R\right)\right]\, .
\end{equation}

\section{Brill--Lindquist data}\label{s4}

Next, we write Brill--Lindquist initial data. The metric has the form (\ref{metric}), but now the conformal factor is given 
by
\begin{equation}
\Phi_{\rm BL}(R,\theta ;z_0)=1+\frac{m}{2}\,\left(\frac{1}{\sqrt{R^2\,\sin^2\theta+(R\,\cos\theta -z_0)^2}}
+\frac{1}{\sqrt{R^2\,\sin^2\theta+(R\,\cos\theta +z_0)^2}}\right)\, ,
\end{equation}
where the two black holes are at positions $\pm z_0$ along the $z$ axis. The parameter $m$ is related to the mass of the 
black holes. 

\subsection{The quasi-Kinnersley frame}

In this Section we find the quasi-Kinnersley frame for B--L data. Our calculation parallels its Misner data counterpart, so we describe it briefly. 
\begin{eqnarray}
C^R_{\;\;R} &=& -4\,\frac{mR^3}{(R+m)^6}-4m\,\frac{6R^2+mR+m^2}{(R+m)^7}\,P_2(\cos\theta)\,z_0^2+O(z_0^4)\\
C^{\theta}_{\;\;\theta} &=& 2\,\frac{mR^3}{(R+m)^6}+2m\,\frac{(7R^2+3mR+2m^2)\,\,P_2(\cos\theta)-(R+m)^2}{(R+m)^7}\,z_0^2+O(z_0^4)\\
C^{\phi}_{\;\;\phi} &=& 2\,\frac{mR^3}{(R+m)^6}+2m\,\frac{R(5R-m)\,\,P_2(\cos\theta)+(R+m)^2}{(R+m)^7}\,z_0^2+O(z_0^4)\\
C^{\theta}_{\;R} &=& \frac{1}{R^2}\,C^R_{\;\;\theta} \nonumber \\
&=& -6m\,\frac{m+4R}{R(R+m)^6}\,\sin\theta\,\cos\theta\,z_0^2+O(z_0^4)
\end{eqnarray}

The eigenvalues of $C^i_{\;j}$ are 
$$\lambda_1=C^R_{\;R}+O(z_0^4)$$
$$\lambda_2=C^{\theta}_{\;\theta}+O(z_0^4)$$
$$\lambda_3=C^{\phi}_{\;\phi}$$
so that the eigenvalue that corresponds to the quasi-Kinnersley frame is $\lambda_1$, and the corresponding eigenvector is $v_1^{\alpha}=a\,(\,\delta^{\alpha}_R+X_{BL}\,\delta^{\alpha}_{\theta})$, where 
\begin{equation}\label{v_BL}
X_{BL}=\frac{4R+m}{R^4}\,\sin\theta\,\cos\theta\,z_0^2+O(z_0^4)
\end{equation}
and
\begin{equation}
a=\Phi^{-2}(1+R^2X_{BL}^2)^{-1/2}=
\frac{R^2}{(R+m)^2}-2\,\frac{m}{(R+m)^3}\,P_2(\cos\theta)\,z_0^2+O(z_0^4)\, .
\end{equation}
(Notice, that $X_{BL}$ affects $a$ only at $O(z_0^4)$.)

We repeat the calculation for Brill--Lindquist initial data, and find that the BB scalar equals
\begin{equation}\label{xi_bl}
\xi_{\rm BL} = \frac{9\,m^2}{(R+m)^{10}}\,\sin^4\theta\,z_0^4
\,+\,
\frac{6\,m^2}{R^3\,(R+m)^{11}}\,\left[(3\,R^2-14\,m\,R-2\,m^2)\,P_2(\cos\theta)-(6\,R^2+7\,m\,R+m^2)\right]\,\sin^4\theta\,
z_0^6+O(z_0^8)
\end{equation}
and the Coulomb scalar is given by
\begin{equation}
\chi_{\rm BL}=-\frac{2\,m\,R^3}{(R+m)^6}-\frac{2m\,(6\,R^2+R\,m+m^2)}{(R+m)^7}\,P_2(\cos\theta)\,z_0^2+O(z_0^4)\, .
\end{equation}

The Weyl scalars in the quasi-Kinnersley frame are
\begin{eqnarray}
\Psi_0 &=& |c|^{-2}e^{2i\vartheta}\,\frac{3m\,\sin^2\theta}{(R+m)^5}\,z_0^2
+ |c|^{-2}e^{2i\vartheta}\,\frac{m\,\sin^2\theta}{R^3(R+m)^6}\,[(3R^2-14mR-2m^2)\,P_2(\cos\theta)+(6R^2+7mR+m^2)]\, 
z_0^4+O(z_0^6)\\
\Psi_2 &=& -2\,\frac{m\,R^3}{(R+m)^6}\,-2m\,\frac{6R^2+mR+m^2}{(R+m)^7}\, P_2(\cos\theta)\,z_0^2 +O(z_0^4)\\
\Psi_4 &=& |c|^{2}e^{-2i\vartheta}\,\frac{3m\,\sin^2\theta}{(R+m)^5}\,z_0^2 
+  |c|^{2}e^{-2i\vartheta}\frac{m\,\sin^2\theta}{R^3(R+m)^6}\,[(3R^2-14mR-2m^2)\,P_2(\cos\theta)+(6R^2+7mR+m^2)]\, 
z_0^4+O(z_0^6)\, .
\end{eqnarray}

\subsection{The quasi-Kineersley tetrad: Breaking the boost symmetry}
For $z_0=0$, $\rho_S=-R/(R+m)^2$. Requiring that the spin coefficient $\rho$ equals its counterpart in the no-radiation limit, we find that
\begin{equation}
|c|=\frac{1}{\sqrt{2}}\,\frac{R-m}{R+m}\, ,
\end{equation}
so that in the quasi-Kinnersley tetrad
\begin{eqnarray}
|\Psi_0| &=& \left|\,\frac{6m\,\sin^2\theta}{(R+m)^3(R-m)^2}\,z_0^2
+ \frac{2m\,\sin^2\theta}{R^3(R+m)^4(R-m)^2}\,[(3R^2-14mR-2m^2)\,P_2(\cos\theta)+(6R^2+7mR+m^2)]\, z_0^4\right. \nonumber \\
&+&\left. O(z_0^6)\,\right| \\
|\Psi_4| &=& \left|\,\frac{3m(R-m)^2\,\sin^2\theta}{2(R+m)^7}\,z_0^2 
+  \frac{m\,(R-m)^2\sin^2\theta}{2R^3(R+m)^8}\,[(3R^2-14mR-2m^2)\,P_2(\cos\theta)+(6R^2+7mR+m^2)]\, z_0^4\right. \nonumber \\
&+&\left. O(z_0^6)\,\right|\, .
\end{eqnarray}
Again, we could also determine the boost degree of freedom using the combination $\Psi_2/\rho^3$. 

Finally, the contribution to the energy flux in gravitational waves at the moment of time symmetry from that moment in time is given by
\begin{eqnarray}
{\cal F}_0 &:=& \int R^2|\Psi_0|^2\,d\Omega\nonumber \\
&=& \frac{3\cdot 2^7\pi\, m^2}{5\, R^2(R+m)^2(R-m)^4}\,z_0^4
-\frac{3\cdot 2^8\pi\, m^2(16R^2+11mR+m^2)}{5\cdot 7R^5(R+m)^3(R-m)^4}\,z_0^6+O(z_0^{8})\\
{\cal F}_4 &:=& \int R^2|\Psi_4|^2\,d\Omega\nonumber \\
&=& \frac{3\cdot 2^3\pi\,m^2(R-m)^4}{5R^2(R+m)^{10}}\,z_0^4
-\frac{3\cdot 2^4\pi\,m^2(R-m)^4(16R^2+11mR+m^2)}{5\cdot 7R^5(R+m)^{11}}\,z_0^6+O(z_0^{8})
\end{eqnarray}

\subsection{The quasi-Kineersley tetrad: Breaking the spin symmetry}
The spin coefficients
\begin{equation}
\varpi = -\sqrt{2}\,i\,\sin\theta\,\cos\theta\frac{e^{-i\vartheta}}{R^2(R+m)}\,z_0^2+O(z_0^4)
\end{equation}
and
\begin{equation}
\tau = -\sqrt{2}\,i\,\sin\theta\,\cos\theta\frac{e^{i\vartheta}}{R^2(R+m)}\,z_0^2+O(z_0^4)\, 
\end{equation}
so that again we can find the ratio (whose limit exists also on the equatorial plane and at the poles) to be
\begin{equation}
\frac{\varpi}{\tau}=e^{-2i\vartheta}\, .
\end{equation}
Doing a spin rotation with, as above,  parameter $\varphi=-\vartheta$ we can set the ratio $\varpi/\tau$ to equal unity as the gauge fixing that gives only $h_+$ gravitational waves, and does not excite the $h_{\times}$ polarization state. In this gauge, the Weyl scalars are given by
\begin{eqnarray}
\Psi_0 &=& \frac{6m\,\sin^2\theta}{(R+m)^3(R-m)^2}\,z_0^2 \nonumber \\
&+& \frac{2m\,\sin^2\theta}{R^3(R+m)^4(R-m)^2}\,[(3R^2-14mR-2m^2)\,P_2(\cos\theta)+(6R^2+7mR+m^2)]\, 
z_0^4+O(z_0^6)\\
\Psi_4 &=& \frac{3m(R-m)^2\,\sin^2\theta}{2(R+m)^7}\,z_0^2 \nonumber \\
&+&  \frac{m\,(R-m)^2\sin^2\theta}{2R^3(R+m)^8}\,[(3R^2-14mR-2m^2)\,P_2(\cos\theta)+(6R^2+7mR+m^2)]\, 
z_0^4+O(z_0^6)\, .
\end{eqnarray}

\section{Comparison of Misner and Brill-Lindquist data}\label{s5}

We cannot yet compare our results for the Weyl scalars $\Psi_0$ and $\Psi_4$, or even our results for the BB scalar for the two cases: the expansion parameter $z_0$ for the case of 
Brill--Lindquist initial data is different from the proper distance $L$ we used in the Misner case; we 
need to relate $z_0$ to the proper distance between the two holes. One way to do this is to find the proper distance between the apparent horizons of the two holes. Alternatively, we require that the Coulomb curvature is the same for both initial data sets (to quadrupole order). Comparing 
$\chi_{\rm M}$ and $\chi_{\rm BL}$ to second order, we find that the relations between the parameters of the two sets are: 
\begin{equation}\label{condition}
M=2m\;\;\;\;\;\; {\rm and} \;\;\;\;\;\; \frac{\zeta (3)}{2\,\pi^6}\, L^3= m\,z_0^2\, .
\end{equation}
Using these two relations, we re-write
\begin{equation}\label{xi_m1}
\xi_{\rm M} = \frac{9\,m^2}{(R+m)^{10}}\,\sin^4\theta\,z_0^4
\,+\,
5\cdot 2^{5/3}\cdot 3\,\frac{\zeta (5)}{\zeta^{5/3}(3)}\,\frac{m^{8/3}\,(7P_2(\cos\theta)+2)}{R^2\,(R+m)^{10}}
\,\sin^4\theta\,z_0^{16/3}+O(z_0^{6})\, ,
\end{equation}
and
\begin{eqnarray}\label{xi_bl1}
\xi_{\rm BL} &=& \frac{2^8\cdot 3^2\,\zeta^2(3)}{\pi^{12}\,(M+2R)^{10}}\,\sin^4\theta\,L^6 \nonumber \\ 
&+&
\frac{2^8\cdot 3\,\zeta^3 (3)}{\pi^{18}\,M\,R^3\,(M+2R)^{11}}\,[2(6\,R^2-14\,MR-M^2)\,P_2(\cos\theta)
-(24\,R^2+14\,MR+M^2)]\,\sin^4\theta\,L^9
\,+\,O(L^{12})\, .
\end{eqnarray}

We see, that the difference between the BB scalars for the two cases appears at $O(L^8)$, and arises from the product of the $O(L^3)$ of one Weyl scalar with the term at $O(L^5)$ of the other. One may therefore expect the energy flux in gravitational waves of the two cases to also differ at $O(L^8)$. However, we find that they differ at $O(L^{10})$. The reason is that the contribution of the $O(L^8)$ term to the energy flux in gravitational waves vanishes after integration over the sphere. The next term, at $O(L^9)$, arising from the product of the $O(L^3)$ and $O(L^6)$ terms, is equal for both initial data sets. At $O(L^{10})$, which has contributions from both the product of the two $O(L^5)$ terms and the product of the $O(L^3)$ and $O(L^7)$ terms, we see the first deviation of Misner data from Brill--Lindquist data: Whereas these terms are non-zero for the Misner case, they vanish for the Brill--Lindquist case as the $O(L^5)$ and $O(L^7)$ terms are absent from the Weyl scalars for Brill--Lindquist initial data. Because the energy flux in gravitational waves is at $O(L^6)$---arising from the product of the $O(L^3)$ terms---the energy flux in the Misner case deviates from its Brill--Lindquist counterpart at a relative order of $O(L^4)$. Specifically, the relation of the energy fluxes in gravitational waves is given by 
\begin{equation}
{\cal F}^{\rm Misner} = {\cal F}^{\rm BL}\,\left[\,1+\frac{25}{3\pi^8}\,\frac{\zeta^2(5)}{\zeta^2(3)}\,
\left(\frac{L}{R}\right)^4+O(L^5)\right]\, .
\end{equation}
Note, that this relation holds separately for the flux in $\Psi_0$ and the flux in $\Psi_4$, and also for the total flux. 

Notice, that we can compare directly the radiation content of Misner data [Eqs.~(\ref{xi_m}) and (\ref{xi_m1})] and Brill--Lindquist data [Eqs.~(\ref{xi_bl}) and (\ref{xi_bl1})] using the BB scalar. We first notice that to $O(L^6)$ (or, equivalently, to $O(z_0^4)$), the BB scalars are the same for both initial data sets. That is, the quadrupole radiations are the same. The reason for that is that our condition (\ref{condition}) for the equality of the Coulomb curvatures is {\em identical} to the condition that the quadrupole amplitudes of the perturbations are equal, $(z_0/M)=2\sqrt{\kappa_2(\mu_0)}$ 
\cite{abrahams}. Namely, by setting the Coulomb curvatures equal, we also set the quadrupole contributions to the BB scalar equal. We only see a difference at the next order: $\xi_{\rm BL}$ has a contribution at $O(z_0^6)$ (or, equivalently, at $O(L^9)$), whereas $\xi_{\rm M}$ has a contribution at $O(z_0^{16/3})$ (or, equivalently, at $O(L^8)$). This implies that at small 
black-hole separations, or $L\ll M$, Brill--Lindquist initial data contain {\em less} gravitational radiation than Misner initial data. The same conclusion was also obtained by Abrahams and Price 
\cite{abrahams}, who found that the Brill--Lindquist solution has a relatively smaller contribution because of higher multipole moments, as its geometry is more quadrupole dominated. Notably, because the difference between the two initial data sets enters at a lower order for the BB scalar than for the energy flux in gravitational waves, it would be easier to observe this difference numerically for the former than for the latter. The BB scalar is therefore a sensitive tool. Note, notwithstanding, that finding the BB scalar is not required for finding the Weyl scalars $\Psi_4$ and $\Psi_0$, which is the objective. As we show in this paper, for the spacetimes we study here (which are both magnetic free) we can find the Weyl scalars.

\section*{Acknowledgments}
The author is indebted to John Baker, Chris Beetle and Gaurav Khanna for invaluable discussions, and to the Aspen Center for Physics for hospitality. This work was supported in part by NASA EPSCoR grant No.~NCC5--580.


\begin{thebibliography}{99}

\bibitem{baumgarte-shapiro} For a recent review see, e.g., T.W.~Baumgarte and S.L.~Shapiro, Phys.~Rept.~{\bf 376}, 41 (2003) 
and references cited therein.
\bibitem{beetle-burko} C.~Beetle and L.M.~Burko, Phys.~Rev.~Lett.~{\bf 89}, 271101 (2002). 
\bibitem{p1} C.~Beetle, M.~Bruni, L.M.~Burko, and A.~Nerozzi, Phys.~Rev.~D {\bf 72}, 024013 (2005).
\bibitem{p2} A.~Nerozzi, C.~Beetle, M.~Bruni, L.M.~Burko, and D.~Pollney, Phys.~Rev.~D {\bf 72}, 024014 (2005).
\bibitem{p3} L.M.~Burko, T.W.~Baumgarte, and C.~Beetle, Phys.~Rev.~D {\bf 73}, 024002 (2006).
\bibitem{p4} A.~Nerozzi, M.~Bruni, V.~Re, and L.M.~Burko, Phys.~Rev.~ D {\bf 73}, 044020 (2006). 
\bibitem{pretorius05} F.~Pretorius, Phys.~Rev.~Lett.~{\bf 95}, 121101 (2005); M.~Campanelli, C.~O.~Lousto, and Y.~Zlochower, Phys.~Rev. D {\bf 73}, 061501(R) (2006); J.~G.~Baker, J.~Centrella, D.-I.~Choi, M.~Koppitz, and J.~van Meter, Phys.~Rev.~Lett.~{\bf 96}, 111102 (2006). 
\bibitem{cook} G.B.~Cook, Living Rev.~Rel.~{\bf 5}, 1 (2000); H.P.~Pfeiffer, G.B.~Cook, and S.A.~Teukolsky, Phys.~Rev.~D {\bf 66}, 024047 (2002). 
\bibitem{nerozzi} A. Nerozzi, M. Bruni, L.M. Burko and V. Re, in {\em Proceeding of the Albert Einstein Century International Conference,} AIP Conference Proceedings {\bf 861}, J.-M. Alimi and A. F\"{u}zfa  (eds.) (American Institute of Physics, 2006), p.p. 702--707 [gr-qc/0607066].
\bibitem{misner} C.W.~Misner, Phys.~Rev.~{\bf 118}, 1110 (1960).
\bibitem{brill-lindquist} D.R.~Brill and R.W.~Lindquist, Phys.~Rev.~{\bf 131}, 471 (1964).
\bibitem{anninos} P.~Anninos, R.H.~Price, J.~Pullin, E.~Seidel, and W.-M.~Suen, Phys.~Rev.~D {\bf 52}, 4462 (1995).
\bibitem{abrahams} A.M.~Abrahams and R.H.~Price, Phys.~Rev.~D {\bf 53}, 1972 (1996).
\bibitem{p6} C.~Beetle and L.M.~Burko, in preparation.
\bibitem{berti-06} E.~Berti, V.~Cardoso, and C.M.~Will, Phys.~Rev.~D {\bf 73}, 064030 (2006) and references cited therein.
\bibitem{bc00} J.~Baker and M.~Campanelli, Phys.~Rev.~D {\bf 62}, 127501 (2000).
\end{thebibliography}
\end{document}